\documentclass[]{interact}
\usepackage{bm} 
\usepackage{lipsum}
\usepackage{graphicx}
\usepackage[separate-uncertainty=true, multi-part-units = single, product-units=single]{siunitx}
\usepackage{upgreek}
\usepackage{epstopdf}% To incorporate .eps illustrations using PDFLaTeX, etc.
\usepackage{subfigure}% Support for small, `sub' figures and tables
\usepackage[numbers,sort&compress,merge]{natbib}% Citation support using natbib.sty
\usepackage{lmodern}
\usepackage{textcomp}

\begin{document}
\title{Quantitative simulation of a magneto-optical trap operating near the photon recoil limit}
\author{\name{Ryan K. Hanley, Paul Huillery, Niamh C. Keegan, Alistair D. Bounds, R.~Faoro, Matthew P. A. Jones\thanks{CONTACT: Matthew P. A. Jones, Email: m.p.a.jones@durham.ac.uk}}
\affil{Joint Quantum Centre (JQC) Durham-Newcastle, Department of Physics, Durham University, South Road, Durham, DH1 3LE, United Kingdom}}
\date{\today}

\maketitle

\begin{abstract}
We present a quantitative model for magneto-optical traps operating on narrow transitions, where the transition linewidth and the recoil shift are comparable. We combine a quantum treatment of the light scattering process with a Monte-Carlo simulation of the atomic motion. By comparing our model to an experiment operating on the $5\rm{s}^2~^1\rm{S}_0 \rightarrow 5\rm{s}5\rm{p}~^3\rm{P}_1$ transition in strontium, we show that it  quantitatively reproduces the cloud size, position, temperature and dynamics over a wide range of operating conditions, without any adjustable parameters. We also present an extension of the model that quantitatively reproduces the transfer of atoms into a far off-resonance dipole trap (FORT), highlighting its use as a tool for optimising complex cold atom experiments. 
\end{abstract}

\section{Introduction}
The advent of laser cooling and trapping \cite{Letokhov1976,Balykin1979,Letokhov1979,Wineland1979} was a revolutionary advance leading to a plethora of ultra-cold atomic experiments. The magneto-optical trap \cite{Raab1987} (MOT) is the workhorse for all experiments with cold neutral atoms and has been extended to molecules in recent years  \cite{Shuman2010,Hummon2013,Zhelyazkova2014,Hemmerling2016}. While quantitative theories of laser cooling in the Doppler and sub-Doppler regimes have existed for many years \cite{Foot2004}, a quantitative model for the MOT is more challenging. The difficulty arises due to the complex three-dimensional polarized light field in the presence of a magnetic quadrupole field, and the effects of optical pumping. In MOTs operating on strong transitions, the re-scattering and absorption of light must also be taken into consideration \cite{Walker1990}. Nonetheless, as the technique is extended to more complex systems such as molecules, there is a considerable interest in models with the ability to quantitatively predict MOT properties.  

One approach typically used to simulate MOT dynamics is to make simplifying assumptions about the atomic system and perform a Monte Carlo \cite{James1980} integration of the classical equations of motion. This assumes that the atoms experience an average force from the laser beams. Wohlleben \cite{Wohlleben2001} and Chaudhuri \cite{Chaudhuri2006} have used this method to accurately simulate the atomic trajectories of rubidium atoms in a 2D$^+$ MOT. This method has also been used to simulate loading into optical traps \cite{Mu2010}. For more complex systems where optical pumping must be included, such as molecular MOTs, this model breaks down. A more accurate model is produced using the optical Bloch equations \cite{Cohen1998}. By performing an adiabatic elimination of the density matrix coherences, one is left with a series of rate equations. Atutov \cite{Atutov2001} has shown this model to be accurate at modelling a sodium MOT involving optical pumping whilst both Comparat \cite{Comparat2014} and Tarbutt \cite{Tarbutt2015} have utilised this method to study the formation of molecular MOTs. 

Divalent atoms exhibit inter-combination lines that are spin-forbidden by the usual electric dipole selection rules, but which are weakly allowed through state mixing, leading to very narrow transitions. These narrow transitions enable the production of  `narrow-line MOTs' (nMOTs) where the dynamics are limited by photon recoil, leading to sub-$\upmu$K temperatures \cite{Castin1989}. The ability to trap atoms at low temperatures has aided the field of precision spectroscopy as the Doppler broadening due to the motion of atoms is greatly reduced, leading to the development of atomic clocks operating in the optical domain \cite{Ludlow2015}. The narrow transition also facilitates the creation of high density atom clouds  \cite{Takasu2003, Loftus2004}, since the radiation trapping that limits the density of conventional MOTs is suppressed. This has led to the a variety of studies in the high density regime, such as the study of quantum degenerate gases \cite{Takasu2003,Kraft2009, Stellmer2009, deEscobar2009,Duarte2011}, multiple scattering \cite{Sesko1991,Labeyrie2003,Sadler2017} and Rydberg blockade \cite{Desalvo2016} to name but a few. 

As the transition linewidth in an nMOT is so small, the transition is often power broadened, which prevents the conventional adiabatic elimination of the density matrix coherences used to a rate equation model \cite{Comparat2014,Tarbutt2015}. To address this, we develop a Monte Carlo model which is based upon the steady-state solution of the optical Bloch equations. 

The paper is structured as follows. We initially introduce the the concept of an nMOT in section 2 and subsequently detail the model in section 3. The experimental configuration used to test the model is detailed in section 4. In section 5, we compare the model to experimental data. In section 6, we explore future applications of this model by simulating the loading of atoms into a far off-resonance dipole trap (FORT). We conclude our findings in section 7. 

\section{Narrow-line MOTs (nMOTs)} \label{nMOT}

The experimental configuration for an nMOT is the same as that for a conventional MOT \cite{Raab1987}, with atoms of mass $m$ cooled and confined by a combination of a quadrupole magnetic field and laser beams (wavelength $\lambda$) with the appropriate circular polarization. What makes nMOTs distinctive is the ratio $\eta =\Gamma/\omega_\mathrm{R}$, where  $\Gamma$ is the natural linewidth of the cooling transition, and $\omega_\mathrm{R}=(4 \pi^2 \hbar )/(2 m \lambda^2)$ is the frequency shift due to the atomic recoil following the absorption or emission of a photon. In conventional magneto-optical traps operating on strong dipole-allowed transitions $\eta \gg 1000$. In this regime a single scattering event does not significantly alter the subsequent probability to scatter a photon, and the effects of individual scattering events can be averaged out, leading to the conventional semi-classical theory of Doppler cooling \cite{Foot2004}.

Conversely, in an nMOT $\eta \approx 1$. This condition typically only occurs when cooling on narrow dipole-forbidden transitions. For example, the $\rm^{88}Sr$ $5\rm{s}^2~^1\rm{S}_0 \rightarrow 5\rm{s}5\rm{p}~^3\rm{P}_1$  transition that we consider in this paper has $\eta =1.6$. In this limit individual scattering events significantly alter the subsequent probability of absorption, and the ultimate limit of laser cooling is set by the recoil temperature rather than the Doppler temperature \cite{Castin1989}.
	
Loftus \textit{et. al} \cite{Loftus2004} showed that the behaviour of atoms in an nMOT is governed by the scaled detuning $\delta =|\Delta|/\Gamma^{\prime} \left(S\right)$, where $\Gamma^{\prime} \left(S\right) = \Gamma \sqrt{1+S}$ is the power-broadened linewidth, and $\Delta = \omega-\omega_0$ the laser detuning with $\omega_0$ and $\omega$ the angular frequencies of the atomic transition and the cooling laser respectively. Here the parameter $S=I/I_\mathrm{Sat}$ is the intensity of the cooling light $I$ normalised by the saturation intensity $I_\mathrm{Sat}$. Three regimes can be identified. 

\begin{figure}
\centering
\includegraphics[width=8.6cm]{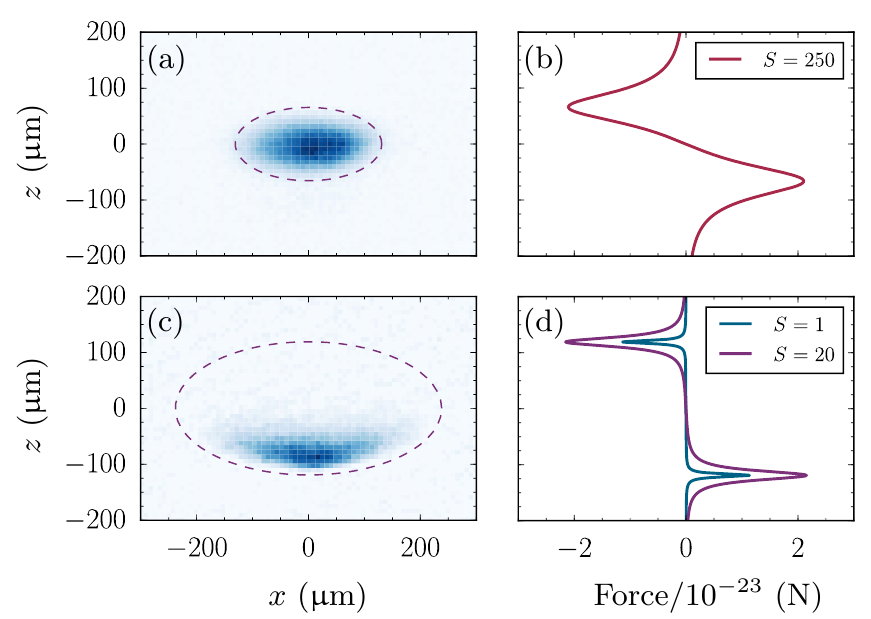}
\caption{Absorption images of the nMOT along with their associated force curves. (a) and (b) correspond to $S=250$ and $\Delta=-2\pi \times \SI{110}{kHz}$ whilst (c) and (d) correspond to $S=20$ and $\Delta=-2\pi \times \SI{200}{kHz}$. The force curve associated with $S=1$ is also shown in (d) for comparison. The dashed ellipse in (a) and (c) shows where $\Delta \omega_z = \Delta$ in the quadrupole field with a gradient of $\SI{8}{G/cm}$.} \label{fig:Figure1}
\end{figure}

The regime that is closest to a conventional MOT occurs when $\delta \approx 1$, and $S>>1$. This is illustrated in figures \ref{fig:Figure1}(a) and (b). Here the power-broadened linewidth dominates and the cloud forms close to the quadrupole centre as atoms are resonant with all three pairs of laser beams. In this ``Doppler'' regime (I) the power-broadened linewidth determines the temperature, and the atoms are comparatively hot.

If $\Delta$ is increased such that $\delta \gg 1$ but $S>1$, then the trap no longer forms at the quadrupole centre, but is displaced to where the Zeeman shift $\Delta \omega_z = \Delta$. The resulting resonance condition forms an elliptical ``shell'' around the quadrupole centre. Since gravity is strong compared to the light-induced forces, the atoms fall under gravity until the resonance condition is met, forming an elliptically-shaped nMOT (shown in figure \ref{fig:Figure1}(c)) where the atoms predominantly interact with the beam that directly opposes gravity. This is seen by the clearly separated force peaks displaced from the quadrupole zero in figure \ref{fig:Figure1} (d). We refer to this as the ``power-broadened regime'' (II)

Finally, the recoil dominated ``quantum'' regime (III) occurs when $\delta \gg 1$, and $S\leq1$. As in the power-broadened regime, the MOT position is determined by $\Delta$ and the magnetic field gradient. However since a photon recoil is sufficient to tune an atom out of resonance with the nMOT beams, recoil effects dominate the scattering behaviour of the atoms. This regime enables the lowest temperatures, ultimately reaching half the photon recoil limit which for $^{88}\rm{Sr}$ is $\SI{460}{nK}$ \cite{Loftus2004}.

\section{Modelling the cloud}
\begin{figure}
\centering
\includegraphics[width=14.4cm]{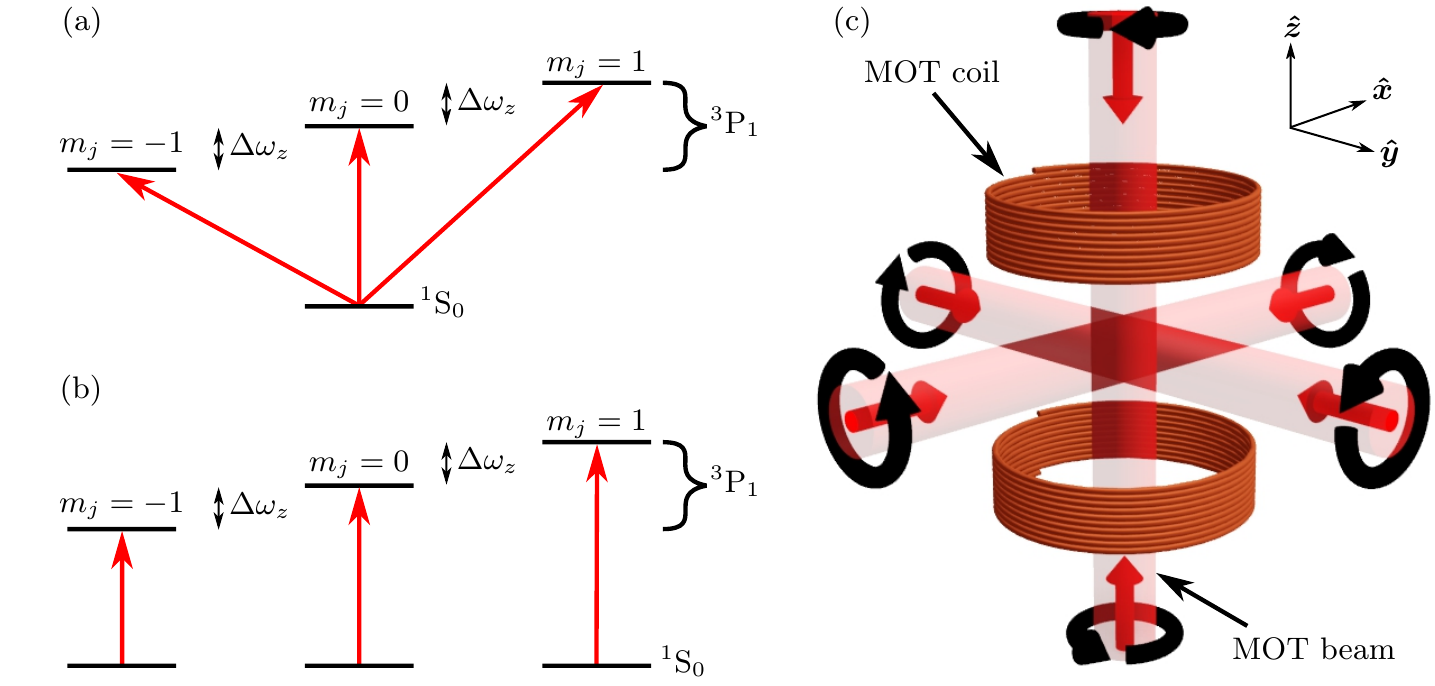}
\caption{(a) Energy level structure of the nMOT. $\Delta \omega_z$ is the Zeeman splitting due to the quadrupole field. (b) Simplified energy level structure used in the simulation. (c) nMOT experimental schematic. The straight red and circular black arrows represent the laser beam propagation direction and helicity respectively.} \label{fig:Figure2}
\end{figure}

In this paper we explicitly consider nMOTs operating on the lowest-lying intercombination lines in divalent atoms such as the alkaline earths and Yb. These $J=0 \rightarrow J=1$ transitions are completely closed\footnote{This is only true for the bosonic isotopes.}, and there is no optical pumping or dark states, as shown in figure \ref{fig:Figure2}(a). Despite this apparent simplicity, it is still very challenging to fully model the interaction of this four-level system with the spatially varying quadrupole field and laser polarization, since one must keep track of complex spatially varying phases between the laser beams that appear in the off-diagonal terms in the atomic density matrix. Therefore, we make a significant simplification and treat each Zeeman transition as an independent two-level system, as shown in figure \ref{fig:Figure2}(b). This amounts to non-conservation of population in the limit of $S\gg 1$ and also neglects Raman-like transitions related to coherences between Zeeman sub-levels. We expect that this is a good approximation in regimes (II) and (III). In these regimes, the atoms fall under gravity until the resonance condition is met and predominantly interact with the laser beam which opposes gravity. The Zeeman splitting between the $m_j$ sublevels is much greater than the transition linewidth, effectively isolating the three Zeeman sublevels, of which the $m_j = -1$ state is most strongly driven due to the helicity of the laser beams.  However we expect the model to fail in the ``Doppler'' regime (I), and we show that this is indeed the case

\subsection{Mathematical formalism}
We simulate the $^{88} \rm{Sr}$ $5\rm{s}^2~^1\rm{S}_0 \rightarrow 5\rm{s}5\rm{p}~^3\rm{P}_1$ nMOT using the conventions and definitions shown in figure \ref{fig:Figure2}(c). The nMOT is constructed from three retro-reflected orthogonal laser beams in the laboratory co-ordinate system $\bm{r} = \left(x,y,z\right)$ where the unit vector directions $\bm{\hat{x}}$, $\bm{\hat{y}}$ and $\bm{\hat{z}}$ are shown in figure \ref{fig:Figure2}(c), and the origin of the coordinate system is taken to be the zero of the quadrupole magnetic field. Each circularly polarized laser beam $i$ has angular frequency $\omega_i$ and wave-vector $\bm{k_i}$ and the helicity of each beam is illustrated in figure \ref{fig:Figure2}(c). The nMOT quadrupole field $\bm{B}$ is defined as ${\bm{B} = \gamma \left(x \bm{\hat{x}} + y \bm{\hat{y}} - 2z \bm{\hat{z}}\right)}$ where $\gamma$ is the gradient of the magnetic field. The magnetic field splits the $^3\rm{P}_1$ state into three Zeeman levels $j$ with splittings $\Delta\omega_z = g\mu_{\rm{B}} \left|\bm{B}\right|/\hbar$ where $g=3/2$ is the g-factor and $\mu_{\rm{B}}$ is the Bohr magneton. 

To reproduce the macroscopic dynamics of the nMOT, the simulation is performed for $\sim 5000$ atoms. Initially, the atoms are uniformly, randomly placed into an user-defined ellipsoid in the laboratory frame with position $\bm{r}$. The atoms are also assigned random velocity vectors $\bm{v}$ taken from a 3D-Boltzmann distribution with a user defined initial temperature. These initial conditions are chosen to be similar to the final nMOT to reduce the processing time\footnote{The simulation is written in Python and makes use of parallelisation packages (psutil) in order to reduce the computation time. A typical running time to simulate an nMOT using a standard desktop computer with an Intel Core i5-4690 Processor with 5000 atoms up to a time of $\SI{15}{ms}$ is $\approx$ 1 hour.}. Typically, an initial temperature of $T=\SI{1}{\upmu K}$ is used. 

The total simulation time $t_{\rm{tot}}$ is broken into time-steps of length $\delta t$. At each time step, the probability that each atom scatters a photon from laser $i$ via a transition $j$ is given by 
\begin{eqnarray}  
P_{ij} &=&  \Gamma_e \rho_{ee}^{ij} \delta t\\ \label{eq:Prob}
&=& \frac{\Gamma_e}{2} \frac{W_jS\delta t}{1+W_jS+4\left(\Delta_i - \bm{k_i} \cdot \bm{v} - \Delta\omega_z^j\right)^2/\Gamma_e^2} ~,
\end{eqnarray}
where $\rho_{ee}^{ij}$ is the steady state excited state population derived from standard two-level optical Bloch equations \cite{Cohen1998} and $\delta t=0.1/\Gamma_e$ such that $P_{ij}\ll 1$. 

The coupling strength $W_j$ is dependent on the local magnetic field and the polarisation of the nMOT laser beam. Due to the spatial inhomogeneity of the magnetic field, $W_j$ must be calculated as a function of position for each laser beam. This is most easily performed by entering a local atomic frame for the calculation. This frame is defined such that the local $z$-axis $\bm{\hat{z}'}$ is directed along the local magnetic field vector. Firstly, the total electric field for each laser beam ${\bm{\mathcal{E}}}$ is defined in the laboratory frame in the spherical basis $\bm{\hat{\epsilon}_q}$ \cite{Brink1993, Auzinsh2010} as  
\begin{equation}
{\bm{\mathcal{E}}} = \sum_q{{\mathcal{E}} ^q \bm{\hat{\epsilon}_q}}~,
\end{equation}
where ${\bm{\mathcal{E}}} = \left({\mathcal{E}}^1, {\mathcal{E}}^0, {\mathcal{E}}^{-1}\right)$ and
\begin{eqnarray}
{\mathcal{E}}^1 &=& -\frac{1}{\sqrt{2}} \left({\mathcal{E}}_x + i{\mathcal{E}}_y\right) \\
{\mathcal{E}}^0 &=& {\mathcal{E}}_z\\
{\mathcal{E}}^{-1} &=& \frac{1}{\sqrt{2}} \left({\mathcal{E}}_x - i{\mathcal{E}}_y\right)~,
\end{eqnarray} 
where ${\mathcal{E}}_{x,y,z}$ is the electric field defined in cartesian coordinates in the laboratory frame. A rotation is then performed to enter the local co-ordinate system of each atom to determine which transitions can be driven along with the associated transition coupling strengths. The rotation matrix is given by 
\begin{eqnarray}
M_q\left(\theta\right) &=& UR\left(\theta\right)U^{\dagger}~,\\
&=& \frac{1}{2}\left( {\begin{array}{ccc}
   1+\cos\left(\theta\right) & -\sqrt{2}\sin\left(\theta\right) & 1-\cos\left(\theta\right) \\
   \sqrt{2}\sin\left(\theta\right) & 2\cos\left(\theta\right) & -\sqrt{2}\sin\left(\theta\right)\\
	1-\cos\left(\theta\right) & \sqrt{2}\sin\left(\theta\right) & 1+\cos\left(\theta\right)\\
  \end{array} } \right)
\end{eqnarray}
where $U$ is the transformation from the cartesian basis to the spherical basis and $R\left(\theta\right)$ is the rotation matrix which maps $\bm{\hat{k}_i}$ onto $\bm{B}$ by an angle $\theta$. This leads to a new polarisation vector ${\bm{\mathcal{E}}^{\prime}} = M_q {\bm{\mathcal{E}}}$ where $W_j=\left|{\mathcal{E}}^{j \prime}\right|^2$.

Once $P_{ij}$ is known, random sampling from a uniform distribution is used to determine whether a scattering event occurs or not. If no scattering event occurs, the atom follows its initial trajectory defined in the laboratory frame as 
\begin{eqnarray} \label{eq:Newton}
\bm{v^{\prime}} &=& \bm{v} + \bm{g} \delta t \\
\bm{r^{\prime}} &=& \bm{r} + \bm{v^{\prime}}\delta t+ \frac{1}{2}\bm{g} \delta t^2~,
\end{eqnarray}
where $\bm{g}=g\left(0,0,-1\right)$ is the acceleration due to gravity and the prime notation represents the final atom position or velocity after a time step $\delta t$. If a scattering event does occur, the atom evolves as 
\begin{equation} \label{eq:Scatter}
\bm{v^{\prime}} = \bm{v} +\frac{\hbar \left|\bm{k_i}\right|}{M}\left(\bm{\hat{k}_i} + \bm{\hat{k}_s}\right) +  \bm{g} \delta t ~,
\end{equation}
where $\bm{k_i}$ is the wavevector of the laser from which the atom initially absorbed a photon with $\bm{\hat{k}_i}$  its associated unit vector, and $\bm{\hat{k}_s}$ is a random unit vector representing the direction of spontaneous emission\footnote{The angular dependence of the spontaneous emission rate is neglected. Although the atoms scatter predominantly on a $\sigma^-$ transition (in regime II and II), which favours emission in the horizontal plane (by a factor of 2), the agreement we observe with the model suggests that this effect is not significant.}.  

During each time step, the atomic positions and velocities are recorded, yielding a complete trajectory of each atom. A simulated absorption image of the nMOT is constructed by histogramming the atomic positions in the $\bm{\hat{x}} - \bm{\hat{z}}$ plane and calculating the column density along $\bm{\hat{y}}$. This is then normalized such that comparisons between theory and experiment can be made. A vertical and horizontal temperature ($T_v$ and $T_h$) is associated with the motion in the $x$ and $z$ directions by fitting a Maxwell-Boltzmann distribution to the vertical and horizontal components of $\bm{v}$. This allows us to obtain the spatial, thermal and temporal dynamics of the atom cloud. 

\section{Experimental Configuration}
The experiment is described in detail elsewhere \cite{Millen2011a,Lochead2012,Boddy2014,Millen2011} and so it is briefly summarised here. Initially, atoms from a strontium oven were slowed using a Zeeman slower before a `blue-MOT' was formed on the $5\rm{s}^2~^1\rm{S}_0 \rightarrow 5\rm{s}5\rm{p}~^1\rm{P}_1$ transition. Atoms in the blue-MOT were cooled to a temperature of several mK. After initial cooling, the blue-MOT light was removed and cooling light at $\SI{689}{nm}$ which addresses the $5\rm{s}^2~^1\rm{S}_0 \rightarrow 5\rm{s}5\rm{p}~^3\rm{P}_1$ transition was applied. This light was artificially broadened to match the Doppler-broadened profile of the atoms in the blue-MOT. After sufficient cooling, the broadening of the $\SI{689}{nm}$ light was removed, leaving single-frequency light, and a cold, dense nMOT. The nMOT was imaged using absorption imaging on the $5\rm{s}^2~^1\rm{S}_0 \rightarrow 5\rm{s}5\rm{p}~^1\rm{P}_1$ transition. Images are captured on a Pixelfly QE camera with a post-magnification pixel-size of $\SI{8}{\upmu m}$. In order to measure the temperature, the cooling light and quadrupole field were switched off, and the atomic cloud was imaged after a variable time-of-flight. By fitting the variation of the cloud width with expansion time we determined the temperature in the $x$ and $z$ directions in the conventional way.

We are able to achieve a range of nMOT sizes, densities and temperatures by varying the nMOT beam detuning, power, and the initial loading rate. We are able to achieve nMOTs ranging in size from $1/e^{2}$ radii of $30~{\rm{to}}~\SI{300}{\upmu m}$, densities up to $\SI{1e{12}}{cm^{-3}}$ and temperatures as low as $\SI{460}{nK}$. The vertical magnetic field gradient is held at $\SI{8}{G cm^{-1}}$ during the final stage of the nMOT. 

\section{Testing the Model} \label{ModelTest}
\begin{figure}
\centering
\includegraphics[width=14.4cm]{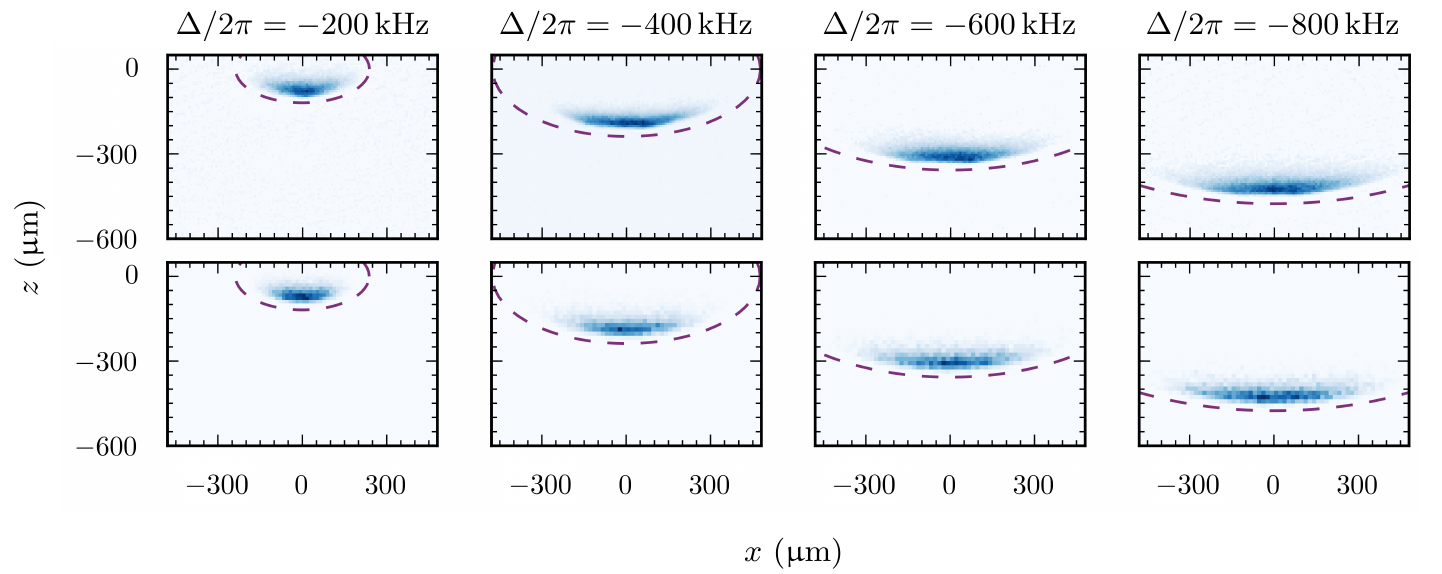}
\caption{Experimental (top row) and theoretical (bottom row) absorption image as a function of nMOT beam detuning with $S=9$. The dashed purple line shows the resonance condition where $\Delta = \Delta \omega_z$.} \label{fig:Figure3}
\end{figure}

As discussed in section \ref{nMOT}, the properties of an nMOT are significantly dependent on $\Delta$ and $S$. This strong parameter dependence allows us to test the accuracy of the model in a wide variety of nMOT regimes. Firstly, we test the model operating in regime (II) where the width and position of the nMOT are strongly dependent on $\Delta$. The top row of figure \ref{fig:Figure3} shows experimental absorption images of the nMOT at four different values of $\Delta$. It is clear that the MOT `sags' under gravity and forms at lower positions as $\Delta$ is decreased. The lower row of figure \ref{fig:Figure3} shows the theoretical absorption images obtained from the simulation. We qualitatively observe excellent agreement in position and shape of the nMOT in the absence of fitting parameters. 

\begin{figure}
\centering
\includegraphics[width=8.6cm]{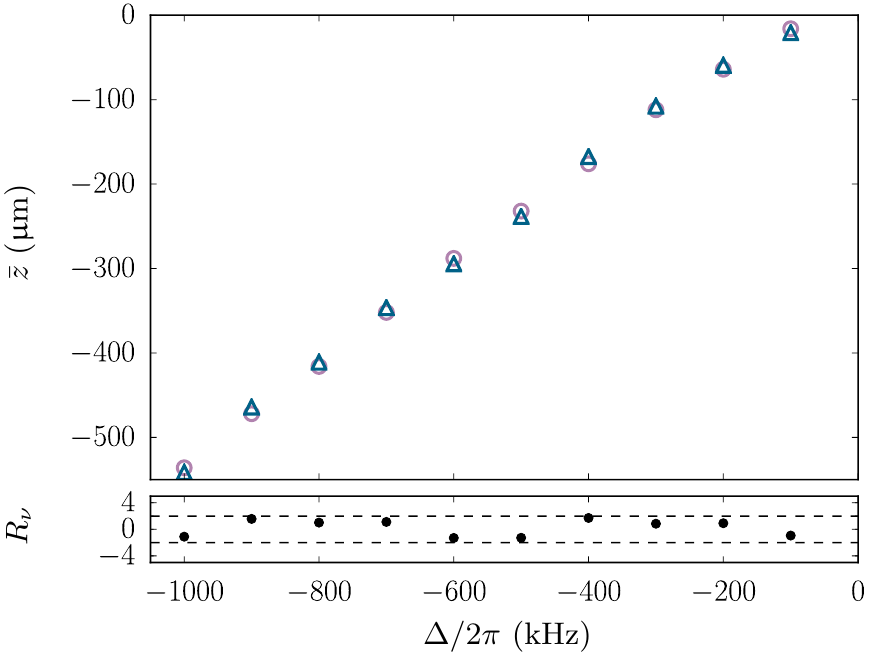}
\caption{Experimental (blue triangles) and theoretical (purple circles) vertical position of the nMOT as a function of nMOT beam detuning. The lower plots show the normalised residuals between theory and experiment, normalised to the error on the experimental data. The dashed black lines show $R_{\nu} = \SI{\pm 2}{}$.} \label{fig:Figure4}
\end{figure}

\begin{figure}
\centering
\includegraphics[width=14cm]{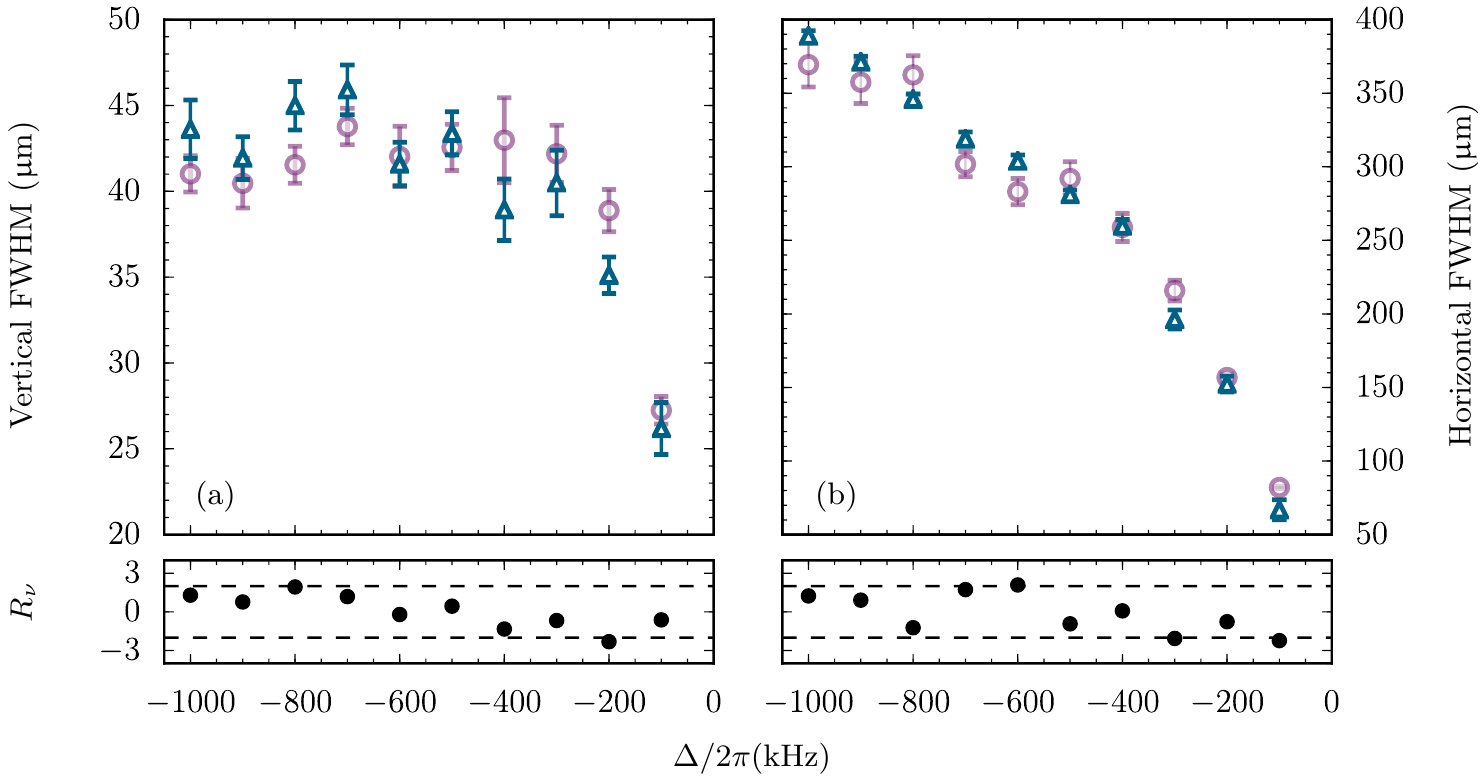}
\caption{Experimental (blue triangles) and theoretical (purple circles) vertical (a) and horizontal (b) FWHM of the nMOT as a function of nMOT beam detuning. The lower plots show the normalised residuals between theory and experiment, normalised to the error on the experimental data. The dashed black lines show $R_{\nu} = \SI{\pm 2}{}$.} \label{fig:Figure5}
\end{figure}

In order to quantitatively compare the model to the experimental data, the mean vertical position $\bar{z}$ and full width at half maximum (FWHM) of the nMOT are extracted numerically (without fitting) from the experimental and theoretical data. The results are plotted as a function of $\Delta$ in figures \ref{fig:Figure4} and \ref{fig:Figure5} respectively. The normalised residuals $R_{\nu}$ \cite{Hughes2010} are shown below each figure. The vertical FWHM saturates as a function of $\Delta$ as the width is determined by the temperature of the atoms. The horizontal FWHM on the other hand continually increases as the radius of the resonant ellipse is proportional to $\Delta$. We observe excellent agreement between experiment and theory with no adjustable parameters. Although the nMOT position is largely determined by the resonance condition, the agreement that we observe indicates that our model also successfully predicts the offset that results from the interplay between the cooling and trapping forces and gravity.

A more stringent test of the model is provided by the temperature. Unlike the position, which is largely determined by the resonance condition, the nMOT temperature is strongly dependent on the intensity of the cooling beams. As $S$ and hence the normalised detuning $\delta$ varies, the nMOT crosses between the different regimes identified in section 2.

\begin{figure}
\centering
\includegraphics[width=8.6cm]{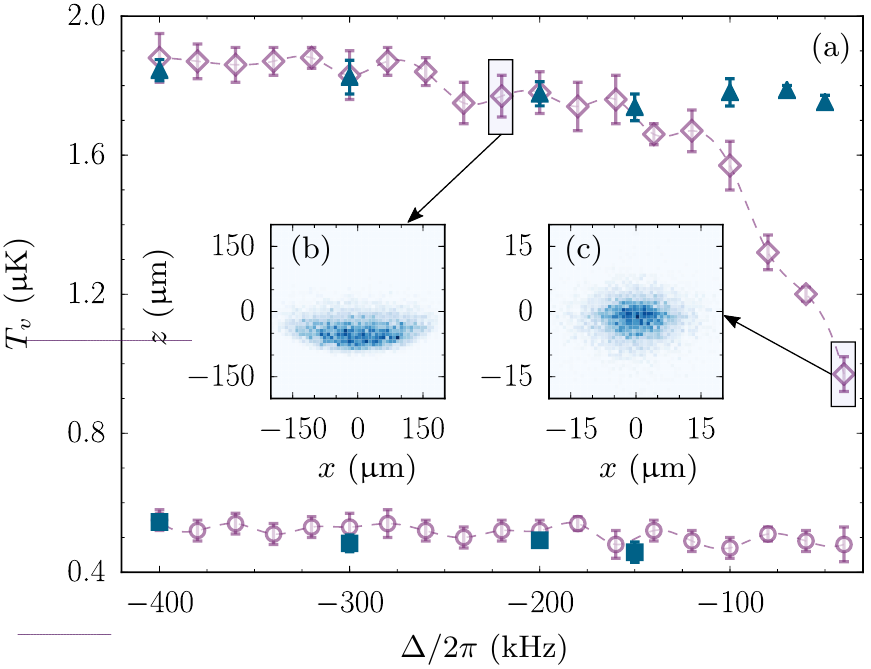}
\caption{(a) The blue squares and purple circles represent the measured and simulated nMOT temperatures for $S=1.9$. The blue triangles and purple diamonds represent the measured and simulated nMOT temperatures for $S=60$. (b) and (c) are the theoretical absorption images for nMOT beam detunings of $- 2\pi \times \SI{220}{kHz}$ and $- 2\pi \times \SI{40}{kHz}$ respectively.} \label{fig:Figure6}
\end{figure}

The dependence of the nMOT temperature on $\Delta$ for two different values of $S$ is shown in \ref{fig:Figure6}(a). Firstly we consider an nMOT operating close to the quantum regime with $S=1.9$. The temperature is essentially independent of $\Delta$, since as shown in figure \ref{fig:Figure4} the position of the MOT just tracks the resonance condition, and the number of scattering events each atoms experiences remains largely unchanged. In this regime, our model is again in excellent agreement with the measurements with no adjustable parameters. 

At higher intensity ($S=60$) the nMOT operates in the power-broadened regime (II). As expected the cloud is hotter, and the temperature is also observed to be largely independent of detuning again. The model is in excellent agreement for ${|\Delta|>2 \pi\times\SI{140}{kHz}}$, but begins to deviate significantly from experiment close to resonance. Here, the power-broadened linewidth begins to approach the Zeeman splitting in the excited state. Thus the MOT crosses over into the conventional ``Doppler'' regime (I) where the linewidth is dominant, forming near the quadrupole zero, as shown in figure \ref{fig:Figure6}(c). As a result, our key assumption that the atoms scatter independently on each of the three Zeeman transitions no longer holds, and the model breaks down.

\begin{figure}
\centering
\includegraphics[width=8.6cm]{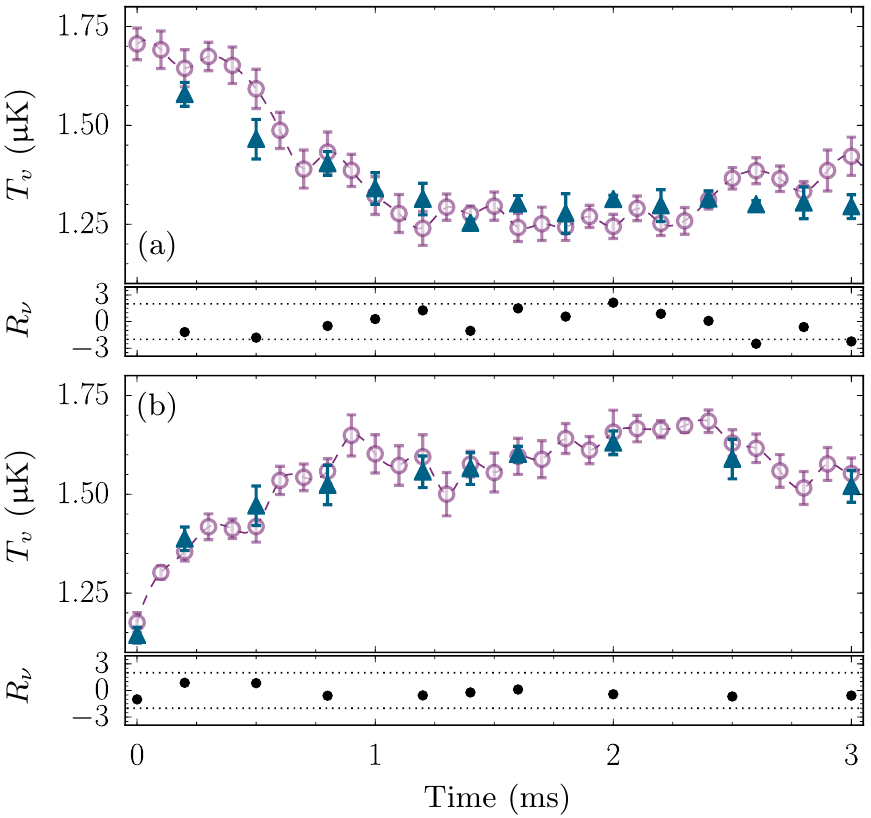}
\caption{Experimental (blue triangles) and theoretical (purple circles) nMOT temperatures as a function of time after a decrease (a) or increase (b) in nMOT laser beam power. The two nMOT beam powers used here were $S = 14$ and 31. The normalised residuals $R_{\nu}$, normalised to the error bars, are shown below each figure. The dashed lines show $R_{\nu} = \pm 2$.} \label{fig:Figure7}. 
\end{figure}

As well as the equilibrium properties, we have also considered whether our model can reproduce the out-of-equilibrium dynamics of the nMOT. To do this, we looked at the response of the temperature to a sudden increase or decrease in the power of the laser beams. Initially, the nMOT was allowed to reach equilibrium at $S=S_0$. At $t=0$, $S$ is suddenly decreased (increased) to a new value $S^{\prime}$. Experimental measurements of the subsequent cooling (heating) are shown in figure \ref{fig:Figure7}, along with the results of the simulation. The agreement is excellent in both cases. More quantitatively, the reduced chi-squared statistics \cite{Hughes2010}  were  $\chi^2_{\nu} = 0.7$ and 1.8 respectively, illustrating that our technique quantitatively reproduces both the steady state and dynamic properties of the nMOT.

In summary, in the regimes of interest for experiment, where the nMOT is cold and dense, the results in figures 3-7 show that our approach yields highly accurate quantitative predictions for the position, size, temperature and dynamics of the nMOT, requiring knowledge only of the experimental control parameters (intensity, detuning and magnetic field gradient). The model breaks down gradually at high intensity and close to resonance, exhibiting significant deviations only when $\Gamma^\prime(S) \approx \Delta$, as expected.

\section{Dipole trapping}

Motivated by the quantitative agreement between theory and experiment reported in section \ref{ModelTest}, we have extended our model to investigate the loading of atoms into a far off-resonance dipole trap (FORT). Optimising the transfer of atoms into such conservative traps is useful in applications such as optical lattice clocks  and Bose-Einstein condensation. Typically, the optimum parameters are found using an experimental exploration of a large parameter space. The ability to quantitatively model the transfer process would therefore be a useful tool. 

To compare the model to experimental data, we simulated the experiment performed by Ido \textit{et al.} \cite{Ido2000}. Their FORT consisted of two crossed laser beams with $1/{\rm{e}}^2$ radius of $\SI{28}{\upmu m}$, operating at $\SI{800}{nm}$. At this wavelength, the differential AC Stark shift between the ground and excited states is negligible, facilitating the simultaneous use of trapping and Doppler cooling. The FORT was loaded by first forming a single-frequency nMOT operating at $\Delta = \SI{-200}{kHz}$ with a total beam intensity of $S=18$. While the nMOT was running, the FORT beams were then applied for a total time of $\SI{35}{ms}$ before the nMOT beams were switched off. The temperature of the atoms in the FORT was subsequently measured using time-of-flight expansion. 

\begin{figure}
\centering
\includegraphics[width=14.4cm]{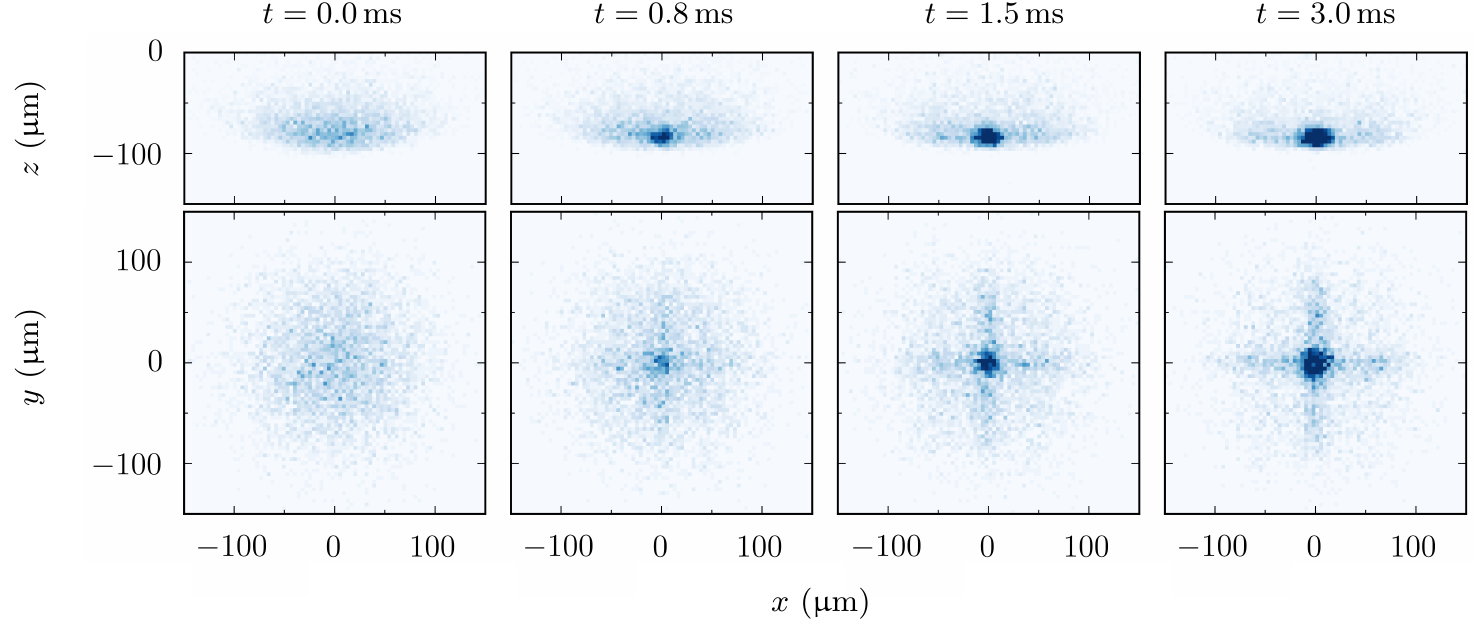}
\caption{Theoretical nMOT dynamics of the simulated crossed FORT with a trap depth of $\SI{5}{\upmu K}$. The top and bottom row shows a theoretical absorption image in the $\bm{\hat{x}} - \bm{\hat{z}}$ and $\bm{\hat{x}} - \bm{\hat{y}}$ plane respectively. All images in each row have the same colour-scale in order to show particle dynamics.} \label{fig:Figure8}
\end{figure}

We included the effect of the FORT beams in our model by neglecting the small differential AC Stark shift of the cooling transition, and considering only the conservative optical dipole force experienced by atoms in the ground state. Thus an extra acceleration is included in the Newtonian dynamics part of the model, given by
\begin{equation}
\bm{a}_{\rm{DT}} = \frac{1}{M} \nabla U\left(x,y,z\right)~,
\end{equation} 
where
\begin{equation}
U\left(x,y,z\right) = U_0 \left(\rm{e}^{-2\left[\left(x-x_0\right)^2+\left(z-z_0\right)^2\right]/w^2} + \rm{e}^{-2\left[\left(y-y_0\right)^2+\left(z-z_0\right)^2\right]/w^2}\right)~.
\end{equation}
$U_0$ is the trap depth, $x_0$, $y_0$ and $z_0$ are linear offsets in the $\bm{\hat{x}}$, $\bm{\hat{y}}$ and $\bm{\hat{z}}$ directions respectively and $w$ is the $1/e^2$ radius of the FORT beams. Figure \ref{fig:Figure8} shows the simulated effect of applying the crossed FORT beams to the nMOT as a function of time. The atoms clearly move into the high intensity region where the FORT beams intersect. We also observe a small number of atoms leaking into each individual FORT beam which is in  qualitative agreement with experimental observations. Ido \textit{et al.} typically capture $\approx \SI{20}{\%}$ of the atoms from the nMOT into the FORT. However, the model predicts a value of approximately double this. We attribute this difference to the lack of collisional losses in the model.

\begin{figure}
\centering
\includegraphics[width=8.6cm]{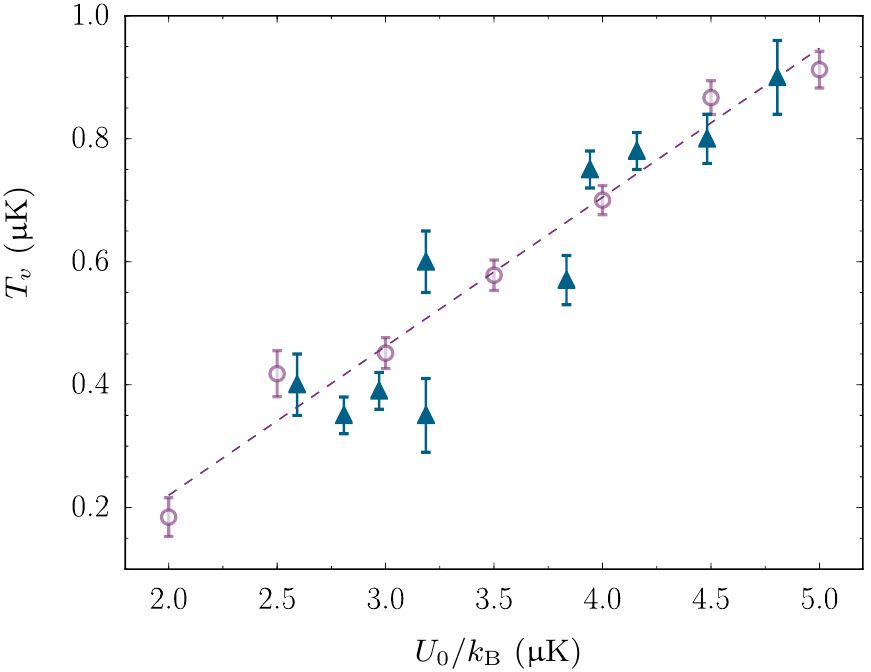}
\caption{Experimental (blue triangles) and theoretical (purple circles) temperatures of the atoms in the crossed FORT as a function of trap depth. The dashed line is a linear fit to the simulated atom temperature. Experimental data taken from \cite{Ido2000}.} \label{fig:Figure9}
\end{figure} 
 
To make a quantitative comparison with experiment, we simulate the temperature of atoms trapped in the FORT as a function of $U_0$. The experimental measurements shown in figure \ref{fig:Figure9} exhibit a linear dependence, with the temperature varying in the range $0.1-0.2\ U_0$. Also shown is the simulated temperature of the atoms captured in the FORT. The error bars on the simulated temperature result from the Maxwell-Boltzmann fit to the velocity distribution. 
%The dashed line is a linear fit to the simulated temperatures, resulting in a gradient of $\SI{0.24 \pm 0.01}{}$. 
We clearly observe excellent agreement between theory and experiment, once again in the absence of any adjustable fitting parameters. Along with the images in figure \ref{fig:Figure8} these results show that it is the interplay between the optical dipole force and laser cooling that sets the temperature, rather than truncation of the velocity distribution or evaporative cooling.

Looking forward, these results illustrate that our model could be a useful tool for optimising the loading parameters of FORTs and optical lattices, eliminating the time-consuming trial and error approach often used to explore the available  parameter space. By adding in a differential AC Stark shift, the model could be easily extended to other trapping wavelengths.

\section{Conclusions}
In summary, we have constructed a semi-classical Monte Carlo simulation in order to model the dynamics of an nMOT, in particular the $\rm^{88}Sr$ $5\rm{s}^2~^1\rm{S}_0 \rightarrow 5\rm{s}5\rm{p}~^3\rm{P}_1$  nMOT. We observed excellent quantitative agreement between theory and experiment without fitting parameters, replicating the spatial, thermal and temporal dynamics of the system. We have also shown that we can quantitatively produce accurate results which simulate the loading of a crossed FORT from an operational three-dimensional nMOT.

In future, we aim to implement atom-atom interactions into the model. It will therefore be possible to fully simulate loading into an optical trap in high density regimes where atom-atom interactions become significant. This will further mitigate the trial-and-error approach of the best parameters for loading atoms into optical traps from nMOTs. We may also be able to simulate the dynamics of a system with large atom-atom interactions such as those displayed by Rydberg atoms, leading to a greater understanding of strongly interacting many-body systems.

\section*{Acknowledgements}
We would like to thank I. Hughes and J. Keaveney for useful discussions. We would also like to thank T. Ido, H. Katori and D. Boddy for supplying experimental data. Financial support was provided by EPSRC grant EP/J007021/, Royal Society grant RG140546 and EU grants FP7-ICT-2013-612862-HAIRS, H2020-FETPROACT- 2014-640378-RYSQ and H2020-MSCA-IF-2014-660028. The data presented in this paper are available for download at (added in proof).  

\bibliographystyle{tfp}
\bibliography{bibfile}

\end{document}